\documentclass[aps,prb,floatfix,preprint,showpacs]{revtex4}

\usepackage{epsfig}

\begin{document}

\title{Mode coupling and evolution in broken-symmetry plasmas.}
\author{E.~Ya.~Sherman$^{1,2}$}
\author{R.M. Abrarov$^{3}$}
\author{J.~E.~Sipe$^{3}$}

\affiliation{$^1$Departamento de Qu\'{i}mica F\'{i}sica, 
             Universidad del Pa\'{i}s Vasco - Euskal Herriko Unibertsitatea, 
             48080 Bilbao, Spain}
\affiliation{$^2$IKERBASQUE Basque Foundation for Science, Alameda Urquijo 36-5,
                  48011, Bilbao, Bizkaia, Spain }
\affiliation{$^3$Department of Physics and Institute for Optical Sciences,
University of Toronto, 60 St. George Street, Toronto, M5S 1A7, Canada}

\begin{abstract}
The control of nonlinear processes and possible transitions to chaos
in  systems of interacting particles is a fundamental physical problem. 
We propose a new nonuniform solid-state plasma system, produced 
by the optical injection of current in two-dimensional semiconductor 
structures,  where this control can be achieved.
Due to an injected current, the system symmetry
is initially broken. The subsequent nonequilibrium 
dynamics is governed by the spatially varying long-range
Coulomb forces and electron-hole collisions. As a result, 
inhomogeneities in the charge and velocity distributions should
develop rapidly, and lead to 
previously unexpected experimental
consequences. We suggest that the system eventually evolves into 
a behavior similar to chaos. 
\end{abstract}
\pacs{05.45.-a,52.35.Mw,72.20.Jv}
\maketitle

Plasmas are of interest  in subjects as diverse as astrophysics and the design
of quantum solid-state nanostructure devices.\cite{Bonilla05,books,Weber06,Pershin07} 
They exhibit a variety of nonlinear phenomena, even close to
equilibrium, including instabilities and chaotic processes
on different scales. \cite{Porkolab78,Davidson91,Yamada08} 
The development of strong turbulence, characterized
by Porkolab and Chang \cite{Porkolab78} as a "stochastic collection
of nonlinear eigenmodes", is a general, and still puzzling, feature of
plasmas.  The Coulomb interaction between
carriers plays the crucial role in producing such
a collection of coupled modes. Due to the very complex 
dynamics, the ability to control the coupling and evolution of 
nonlinear eigenmodes  is a challenging problem. 

Recent progress in optical phase control
allows the production of plasmas in semiconductors with a well-controlled 
\textit{charge density} and, more importantly, a well-controlled \textit{current density}.
\cite{vanDriel01,Atanasov96,Duc06}
The control  of the initial current density is achieved by the quantum 
interference of a one-photon transition (light frequency 
$2\omega$, with the field phase $\phi_{2\omega}$) and a two-photon 
transition (light frequency $\omega$, with the field phase $\phi_{\omega}$)
across the fundamental band gap $E_g$.
At nonzero  $\Delta\phi\equiv\phi_{2\omega }-2\phi_{\omega}$ the symmetry 
of the injected distribution in momentum space
is broken, and a macroscopic current with a 
speed $U_{0}=v_{e}|\sin\Delta\phi|$ is injected in a direction parallel to
the sample surface. 
The maximum speed of the injected electrons, $v_{e}$, is determined by 
$\omega$ and $E_g$, reaching 10$^3$ km/s for $2\hbar\omega-E_{g}$ about 100 meV.

Studies of nonequilibrium electron processes in semiconductors \cite{Chicago}
show that the entire dynamics is complex even for a uniform electron density. When current is injected, 
the resulting separation of electrons and holes leads to strongly 
nonuniform Coulomb forces. Here we consider situations where these forces 
determine, rather than just perturb, the development of the 
charge and current density patterns that can lead to possible nonlinearities and
instabilities. The system  we study theoretically is a multiple quantum well (MQW)
structure, consisting of up to tens GaAs/Al$_x$Ga$_{1-x}$As periods, each of 
thickness on the order of 15 to 30 nm, grown along the $z$ direction. 
At photon energies where carriers are injected only in the GaAs layers,
the total thickness $w$ of the region, that is the number of periods 
multplied by the period width, where the plasma is produced in typical MQWs can be on the order 
of 0.1 $\mu$m, still considerably less than the spot size of the exciting laser beams
and the light absorprion length; the fact that allows to treat all single quantum wells 
as equivalent electrostatically coupled layers, neglecting the direct motion
of elecrons between the wells.  
The injected carrier densities are typically Gaussian in the two-dimensional
coordinate ${\mathbf r}=(x,y)$, given by 
$N_{e,h}(\mathbf{r,}t=0)=N_{0}\exp \left(-r^{2}/2\Lambda ^{2}\right) $,
($e$ for electrons, $h$ for holes) 
where $\Lambda$ is the spot size, and $N_{0}$ is the maximum total injected two-dimensional 
density for all quantum wells, which is proportional to the
total number of single quantum wells and can be on the order of $10^{13}$ cm$^{-2}$.
$N_0$ is the concentration parameter in our analysis.  
As a result, the three-dimensional density distribution can be modelled \cite{Abrarov07} as uniform 
along the $z$-axis, with:
\begin{equation}
N_{e,h}^{[3D]}(\mathbf{r},z;t)=\frac{1}{w}N_{e,h}(\mathbf{r},t)\theta(z)\theta(w-z).
\end{equation}
For this reason we treat the density and velocity distributions in the $(xy)$ plane only. 

The in-plane electric field depends on
an integral over the charge density $-eN_{c}\left(\mathbf{r},t\right)$,  where
$N_{c}\left(\mathbf{r},t\right)\equiv N_{e}\left( \mathbf{r},t\right) -N_{h}\left( \mathbf{r},t\right),$ 
$e$ is the fundamental charge, and is 
given by 
\begin{equation}
\mathbf{E}(\mathbf{r},t)=-\frac{e}{\epsilon }\int N_{c}(%
\mathbf{r}^{\prime },t)\mathbf{K}_{C}\left( \mathbf{r-r}^{\prime }\right)
d^{2}r^{\prime },  \label{Efield}
\end{equation}
where the model Coulomb kernel 
$\mathbf{K}_{C}\left(\mathbf{d}\right) =\mathbf{d}/\left(
d^{2}+w_{C}^{2}\right)^{3/2}$ takes into account the
width of the system and simplifies the calculations
by avoiding the singularity at $d=0$. Here $\epsilon$ is the background
dielectric constant. The parameter $w_C$ is on the order of structure width, where for $w_C\ll\Lambda$
the results are not sensitive to the kernel behavior at $d\ll\Lambda$.
The field $\mathbf{E}(\mathbf{r},t)$ is very sensitive to 
the details of the carrier dynamics, since even relatively small changes in $N_{e}(\mathbf{r},t)$ can 
strongly modify it. 
For example, even if $N_{e,h}(\mathbf{r},t)$ are taken to be 
slightly separated identical Gaussian 
profiles, Eq.(\ref{Efield}) shows 
that $\mathbf{E}(\mathbf{r},t)$ is strongly nonuniform.
Nonuniformities in the field and in the
velocities and the charge patterns mutually enhance each
other.  This process is our main interest. 

To study the nonlinear dynamics, we employ a hydrodynamic model for the 
dynamics of injected charge currents and densities, 
and include the possibility that an external electric 
field is also present. 
In hydrodynamic models one avoids requiring the details of
distribution functions by constructing approximate, closed sets of equations
involving conserved and slowly-varying quantities such as charge, momentum,
and energy densities. In the effective mass approximation, 
closed equations in the range of parameters we consider
can be obtained for the velocity and density.\cite{Abrarov07} 
For simplicity, we assume that the holes in the injected plasma
are not moving, which does not qualitatively 
influence our results\cite{Abrarov07}
due to a small effective mass ratio of electrons and holes. 
The injection typically occurs on a time scale
of 50-100 fs. We take this as instantaneous, and treat it as preparing our initial state.
Since the timescales of interest are much
shorter than electron-hole recombination times, the dynamics is
governed by the continuity equation
for the electron density and the Euler equation:
\begin{eqnarray}
&&\frac{\partial N_{e}}{\partial t}+\nabla \left( N_{e}\mathbf{u}\right)=0, \nonumber \\ 
&&\frac{\partial \mathbf{u}}{\partial t}
+\left(\mathbf{u}\nabla\right)\mathbf{u}
+\frac{\nabla P}{m_{e}N}
=-\frac{e(\mathbf{E}+\widetilde{\mathbf E})}{m_{e}}-
\frac{\mathbf{u}}{\tau _{\rm eh}}\frac{N_{h}}{{N_{0}}}-
\frac{\mathbf{u}}{\tau_{e}},  \label{dNdudt_partial} 
\end{eqnarray}
where $\widetilde{\mathbf E}$ is a time-dependent external electric
field. Here and below the $\mathbf{r}$ and $t-$arguments are omitted for brevity;
$P$ is the pressure, $m_{e}$ is the electron effective mass, 
the weakly concentration-dependent $\tau_{\rm eh}$
describes momentum-conserving drag\cite{Zhao07} due to the Coulomb forces during
electron-hole collisions,\cite{Hopfel} $\tau_{e}$ is the relaxation time
due to external factors, such as phonons  \cite{Knorr97} and disorder. 
Here we consider the effect of this drag only, 
assuming $\tau_{e}\gg\tau_{\rm eh}$ for a clean sample and
electron energies too low for intense phonon emission.\cite{Abrarov07} The electron-hole
drag and the Coulomb forces, being coordinate-dependent, increase
the inhomogeneity in the charge density. 

To obtain the solution of equations (\ref{Efield}),(\ref{dNdudt_partial})
we use a finite basis set, following the Galerkin
method, and convert Eqs.(\ref{dNdudt_partial}) 
to a system of ordinary differential equations. 
The expansion has the form: 
\begin{equation}
N=\sum_{\overline{n}}^{n_{\max }}N_{\overline{n}}^{e}(t)\Psi _{\overline{n}},%
\text{ \ }u_{i}=\sum_{\overline{n}}^{n_{\max }}u_{\overline{n}}^{i}(t)\Psi
_{\overline{n}}+U_{i}\left( t\right), 
\label{expansion}
\end{equation}
where $i=x,y$ is the Cartesian index. To improve the convergence, we include
known functions of time $U_{i}\left( t\right) $ in the right-hand-side of
Eq.(\ref{expansion}) for velocities. These functions can be obtained by
solving the equations of motion in the rigid-spot approximation \cite
{Sherman06} where the electron puddle moves with uniform velocity 
$\mathbf{u}=(U_x(t),U_y(t))$ while keeping 
its initial Gaussian shape. The initial distribution $N_{0}\exp \left(-r^{2}/2\Lambda ^{2}\right)$
suggests the eigenstates
of a harmonic oscillator $\Psi _{\overline{n}}\left( x,y\right)=\psi _{n_{1}}(x)\psi_{n_{2}}(y)$
as the basis set of the expansions with:
\begin{equation}
\psi _{n}(x) =\frac{1}{\sqrt{\sqrt{\pi }n!2^{n}}}
H_{n}\left( x/\Lambda \right) e^{-x^{2}/2\Lambda ^{2}},
\nonumber
\label{basis}
\end{equation}
where $H_{n}\left( x/\Lambda \right) $ is the Hermite polynomial of the $n$%
th order, and the double-index $\overline{n}\equiv (n_{1},n_{2})$. The 
basis functions satisfy the conditions for norm and derivatives:
\begin{eqnarray}
&&\int_{-\infty }^{\infty }\psi _{n_{2}}(x)\psi _{n_{1}}(x)dx=\Lambda \delta
_{n_{1},n_{2}},  \label{conditions} \\
&&\sqrt{2}\Lambda \psi _{n}^{^{\prime }}(x)=\sqrt{n}\psi _{n-1}(x)-\sqrt{n+1}%
\psi _{n+1}(x). \nonumber 
\end{eqnarray}
In this basis, the matrix elements for the components of the Coulomb integrals 
$\mathcal{C}_{\overline{m};\overline{n}}^{i}$ (\ref{Efield}) are given
by: 
\begin{equation}
\mathcal{C}_{\overline{m};\overline{n}}^{i}=\frac{1}{\Lambda ^{2}}\int \int
\Psi _{\overline{m}}\left( \mathbf{r}\right) K_{C}^{(i)}\left( \mathbf{r}-%
\widetilde{\mathbf{r}}\right) \Psi _{\overline{n}}\left( \widetilde{\mathbf{r%
}}\right) d^{2}\widetilde{r}d^{2}r.  
\label{Coulomb}
\end{equation}
Taking into account (\ref{conditions}), the equations of
motion can be written in the operator form: 
\begin{eqnarray}
&&\frac{dN_{\overline{m}}}{dt}=
\frac{1}{\sqrt{2}\Lambda}
\left\{
\sum_{\overline{n},\overline{k%
}}N_{\overline{n}}\left( u_{\overline{k}}^{x}P_{{2}}
\hat{L}_{12}
P_{{1}}+u_{\overline{k}}^{y}\,P_{{1}}
\hat{L}_{12}
P_{{2}}\right) -U\left( \hat{\ell }_{1}^{\dagger}-\hat{%
\ell }_{1}\right) N_{\overline{m}}
\right\},  \nonumber \\
&&\frac{du_{\overline{m}}^{x}}{dt}=-\frac{e^{2}}{\epsilon
m_{e}}\sum_{\overline{k}}N_{\overline{k}}\mathcal{C}_{\overline{m};\overline{%
k}}^{x}-\sum_{\overline{n},\overline{k}}\frac{u_{\overline{k}}^{x}}{\tau
_{\rm eh}}\frac{N_{\overline{n}}^{h}}{{N_{0}}}P_{{1}}P_{{2}}-\frac{U%
}{\tau _{\rm eh}}\frac{N_{\overline{m}}^{h}}{{N_{0}}}+I_{\overline{m}}\left( 
\frac{e\widetilde{E}}{m_{e}}-\frac{dU}{dt}\right), 
\label{dNdudt}
\end{eqnarray}
where the equation for  $du_{\overline{m}}^{\left(y\right)}/dt$  is similar to the latter.
Here we assume $\widetilde{\mathbf E}$ parallel
to the $x$-axis, and put $U_{y}\equiv 0$ and $U\equiv U_{x}$
for the current injected
along the $x$-axis.
The small terms $\left(\mathbf{u}\nabla\right)\mathbf{u}$ and $\nabla P/(m_{e}N)$
in the Euler equation have been neglected; the justification of this approximation
will be given later in the text.  
We have put:
\begin{eqnarray}
P_{i}\equiv P_{n_{i},k_{i},m_{i}} &=&\int_{-\infty
}^{\infty }\psi _{n_{i}}\left( x\right) \psi _{k_{i}}\left( x\right) \psi
_{m_{i}}\left( x\right) \frac{dx}{\Lambda},  \label{overlap} \nonumber \\
I_{\overline{m}} &=&\int_{-\infty }^{\infty }\psi
_{m_{1}}\left( x\right) \frac{dx}{\Lambda}
\int_{-\infty }^{\infty }\psi _{m_{2}}\left(
x\right) \frac{dx}{\Lambda}.  
\end{eqnarray}
The operator 
$\hat{L}_{12}\equiv \hat{\ell}_{1}^{\dagger}+\hat{\ell }_{2}^{\dagger}-\hat{\ell }_{1}-\hat{\ell
}_{2}$, where the ladder operators  
$\hat{\ell }_{p}$ and $\hat{\ell }_{p}^{\dagger}$ act on the 
corresponding \textit{index}, for example: $\hat{\ell }_{2}P_{i}=\sqrt{k_{i}}P_{n_{i},k_{i}-1,m_{i}}.$
For the problem we consider here, the initial conditions are: 
$N_{\overline{n}}(0)=\sqrt{\pi}N_{0}\delta _{n_{1},0}\delta _{n_{2},0}$ and $u_{\overline{n}}^{i}(0)=0$, 
where $N_{\overline{n}}^{h}$ is nonzero only if $%
n_{1}=n_{2}=0$ and remains constant in time. 
Some of the interesting gross quantities that can be calculated with these equations will 
be analyzed below.

The electron-hole drag makes $%
u_{\overline{m}}$ dependent on all components of the velocity. 
Despite this complication, Eqs.(\ref{dNdudt})
can be solved directly in the case of vanishing long-range Coulomb forces. The resulting charge density 
has the form:
\begin{equation}
\frac{N_{c}(\mathbf{r},t)}{N_e(\mathbf{r},0)}=\frac{U_{0}tx}{\Lambda ^{2}}\exp
\left( -\frac{t}{\tau _{\rm eh}}e^{-r^{2}/2\Lambda ^{2}}\right) .  \nonumber
\end{equation}
The inclusion of long-range Coulomb forces leads to a much more complex dynamics. 
In Eqs.(\ref{dNdudt}) the Coulomb matrices $\mathcal{C}_{\overline{m};\overline{k}}$ 
couple a given velocity component 
$u_{\overline{m}}$ to all density components $N_{\overline{k}}$ allowed by symmetry. In
turn, the density evolution depends on all possible products of components
of velocity and density. Therefore, a perturbation in one component can
cause a growing response in a large number of them.  
The temporal behavior of the system is determined by three
independent time scales: drag-induced $\tau _{\rm eh}$; the plasma period $T_{\rm pl}=2\pi/\Omega_{\rm pl}$, where 
$\Omega_{\mathrm{\rm pl}}$ is the two-dimensional plasma frequency for the Gaussian density distribution,
\cite{Sherman06,Ando82} $\Omega_{\mathrm{\rm pl}}^{2}=\pi^{3/2}N_{0}e^{2}/(4\epsilon m_{e}\Lambda)$;
and the timescale of the external $\widetilde{\mathbf E}(t)$. 
We use the parameter $p\equiv\Omega_{\rm pl}\tau _{\rm eh}$
to characterize the relative effects of the long-range Coulomb forces and drag.

In our simulations we use GaAs $m_{e}=0.067m$,
where $m$ is the free electron mass and the dielectric constant $\epsilon =12$;
we take $N_{0}=10^{13}$ cm$^{-2}$ and $\Lambda =1$ $\mu$m. These parameters result in
a plasma period $T_{\mathrm{\rm pl}}$ close to 0.9 ps, which is considerably larger
than the injection time.  
The parameter $w_{C}$ in the Coulomb kernel is taken as $0.1\Lambda$. 
The basis set includes 32 states for each coordinate, giving a convergence \cite{nmax}
in the time interval of interest $0<t<T_{\mathrm{\rm pl}}/2$. 
We consider different values of $p$ in the experimentally achievable range.\cite{Zhao07}

\vspace{10mm}
\begin{figure}[t!]
\begin{center}
\includegraphics[width=.4\columnwidth]{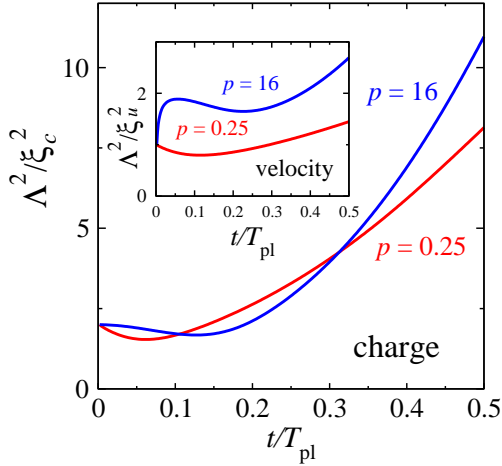}
\end{center}
\par
\vspace{5mm}
\caption{(Color online.) Inhomogeneity parameters of the charge 
density (main plot) and  velocity (inset)
pattern for two regimes of the Coulomb forces. The parameter $p\equiv\Omega_{\rm pl}\tau _{\rm eh}$ is shown near the plots.
$p=16$ corresponds to the extremely weak damping, while for $p=0.25$ the damping is relatively strong.
Here $T_{\rm pl}=0.9$ ps. The functions presented in the plots are universal in the
sense they do not depend on the initial speed of the puddle $U_0$. 
}
\label{fig1:Inhomogeneity}
\end{figure}

To trace the evolution in the inhomogeneity of the
charge density and velocity patterns, we 
study $\xi _{c}$ and $\xi_{u}$, defined to be ratios of gross quantities: 
\begin{eqnarray}
&&\frac{1}{\xi _{c}^{2}(t)}{\int} N_{c}^{2}d^{2}r\equiv\int\left( \nabla
N_{c}\right) ^{2}d^{2}r, \\
&&\frac{1}{\xi _{u}^{2}(t)}\int\left(u_{x}-U\right)^{2}d^{2}r\equiv\int\left(\nabla u_{x}\right)^{2}d^{2}r,
\nonumber
\end{eqnarray}
that serve as characteristic lengths. Taking into account that 
the spatial inhomogeneity (internode distance) of the function $\psi_n(x)$ scales at large $n$ as 
$n^{-1/2}$, the number of harmonics forming the corresponding pattern scales as 
$\Lambda^2/\xi_{c}^{2}(t)$ or $\Lambda^2/\xi_{u}^{2}(t)$ if the distributions
are strongly nonuniform. 
As one can see in Fig.1, both patterns, especially the density, become strongly 
inhomogeneous and the role and the number of the higher
harmonics grows with time. Therefore, we expect that the
spatial scales of the variations in the density and velocity 
will rapidly decrease. Eventually, a hydrodynamic description will fail, as
stochastic behavior develops.\cite{chaos} 
\vspace{10mm}
\begin{figure}[h!]
\begin{center}
\includegraphics[width=0.4\columnwidth]{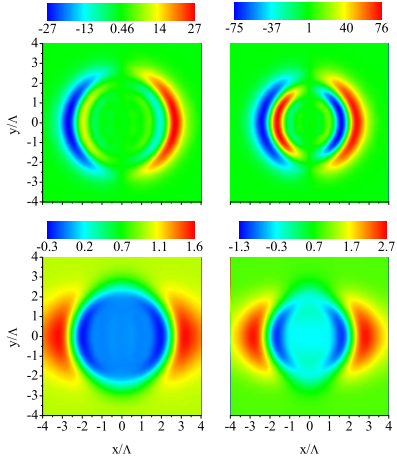}
\end{center}
\par
\vspace{5mm}
\caption{(Color online.) Patterns (in arbitrary, same for both columns, units) 
of charge density $N_{c}(x,y)$ (upper row) and velocity
$u_{x}(x,y)$ (lower row) at $t=T_{\rm pl}/2$. Left column: $p=0.25$, right column: $p=16$.
The density has $N_{c}(x,-y)=-N_{c}(x,y)$ symmetry. In the upper row, larger bow at $x>0$ corresponds to 
$N_{c}(x,y)>0$. The velocity satisfies the condition $u_x(-x,y)=u_x(x,y)$. For $u_x(x,y)$ maximum 
values are achieved at the wings $|x|/\Lambda$ close to 3, $y=0$. Minimum values
are achieved at $|x|/\Lambda$ close to 1, $y=0$.}
\label{fig3:Plots2D}
\end{figure}

The underlying charge density is shown in the upper panel of Fig.2
where we plot the profiles $N_{c}(x,y,T_{\rm pl}/2)$.
The lower panel shows the velocity $u_{x}(x,y,T_{\rm pl}/2)$. The
profiles have a rather complex form, showing that the distributions of both
quantities are strongly inhomogeneous.
\begin{figure}[h!]
\begin{center}
\vspace{30mm}
\includegraphics[width=.4\columnwidth]{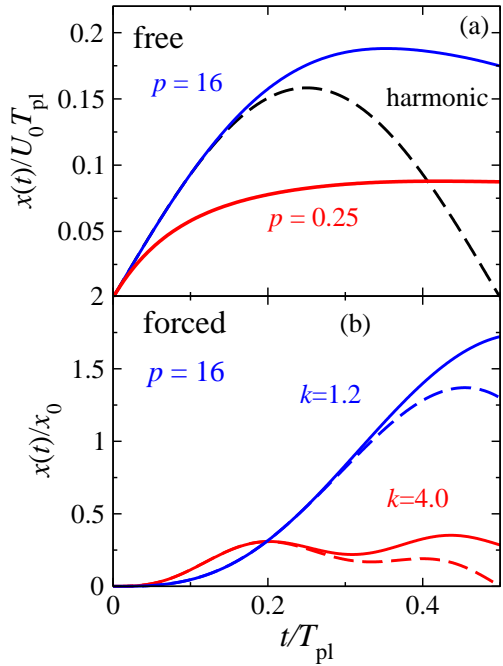}
\end{center}
\par
\vspace{5mm}
\caption{(Color online.) (a) Mean spot displacements for free spot
propagation. Dashed line corresponds to the linear undamped oscillations. The displacement
of the spot is on the order of $U_0T_{\rm pl}/4$, that is 20 nm for
typical $U_0=100$ km/s and assumed here $T_{\rm pl}=0.9$ ps.
(b)  Mean spot displacements for the forced oscillations. 
Dashed lines correspond to the linear oscillator in Eq.(\ref{forced}).
The frequency coefficients are $k=4$ (off-resonance) and $k=1.2$
(close to the resonance).}
\label{fig1:X_av_256}
\end{figure}
We calculate the mean spot displacement:
\begin{equation}
x(t)=\sum_{\overline{n}}\frac{N_{\overline{n}}^{e}}{N_{t}}\int x\Psi _{%
\overline{n}}\left( x,y\right) d^{2}r,  \label{xav}
\end{equation}
where $N_{t}=\pi N_{0}\Lambda^{2}$ is the total number of injected 
electrons. The displacement $x(t)$ has a complex time-dependence,
after initially evolving simply as $U_{0}t$. Even at later times 
$x(t)$ is  proportional to $U_0$ if all other parameters are kept the same. We show in Fig.3 the mean displacement
$x(t)$  defined in Eq.(\ref{xav}) for two different cases presented in Fig.1:
considerably ($p=0.25$) and very weakly 
($p=16$) damped regimes. An astonishing
result is the absence of the plasma oscillations even close to the
clean limit with $p=16$. On the timescale of half of the expected oscillation period 
$T_{\rm pl}$, the spot becomes strongly inhomogeneous with harmonics up to $n_{1},n_{2}\leq 20$
contributing to the results. Therefore, no well-defined
oscillations occur. In all cases considered, the maximum of 
$x(t)\sim U_{0}\min(T_{\rm pl},\tau_{\rm eh})$ is much less than 
$\Lambda$, and therefore the $\nabla{\mathbf u}$ and
$\nabla{P}$-originated terms in the Euler equation can be neglected. 

As another example of this unusual behavior,
we present the results for the clean system ($p=16$) driven by 
an external field $\widetilde{E}(t)=E_{0}\sin (k\Omega_{\rm pl}t)$
for the same initial Gaussian
density distribution as above, but with no current injection ($U_0=0$). 
Here the inhomogeneity develops more slowly than if current were 
injected, since $x(t)$
increases as $t^3$ rather than as $t$ at the initial stage of the process. 
Nonetheless, the $x(t)$ is considerably different from the expected for a linear
oscillator:
\begin{equation}
x_{\rm lo}(t)=\frac{x_{0}}{1-k^{2}}
\left(\sin k\Omega_{\rm pl}t-k\sin\Omega_{\rm pl}t\right),
\label{forced}
\end{equation}
with $x_{0}={eE_{0}}/m_{e}\Omega_{\rm pl}^{2}$,
due to the fact that the excitation of the higher Hermite-Gaussian modes strongly influences
the response to the external field, as shown in Fig. (\ref{fig1:X_av_256}b). 
For a system driven close to resonance
($k=1.2$), the difference between the full and linear oscillator behavior
is less than for $k=4$, since near resonance the uniform external
force is more important than the interactions.  

To conclude, the macroscopic dynamics of optically injected currents in 
clean semiconductor multiple quantum wells is strongly inhomogeneous and nonlinear,
due to the nonuniform long-range Coulomb forces that develop. 
These forces arise following the initial breaking of the 
symmetry by the injected electron puddle velocity ${\bf U}_{0}$,
which leads to a separation of electrons
and holes that produces the nonuniform macroscopic Coulomb interaction. 
Due to the coupling of the Hermite-Gaussian modes through
conservation of charge, the charge density becomes nonuniform on progressively smaller spatial
scales. In contrast to what might be expected, it does not show well-defined 
plasma oscillations.  The complex charge and current density 
patterns develop on a time scale on the order of a quarter
of the plasma oscillation period characteristic of the given carrier density and
puddle size. The length scales characterizing the spatial inhomogeneities 
in density and velocity decrease rapidly, and, in the terminology of Porkolab and Chang [\onlinecite{Porkolab78}], 
a turbulence regime will likely develop. These systems will provide a new laboratory 
example of plasmas with controlled non-linear behavior, and likely a transition
to a stochastic regime.

\textit{Acknowledgement}. This research was funded by the University of the Basque Country (grant GIU07/40),
Natural Sciences and Engineering Research Council of Canada
(NSERC) and the Ontario Centres of Excellence (OCE).


\begin{thebibliography}{99}
\bibitem{nmax} The physical observables were calculated by summing the 
               contributions of the states up to $n=25$ to avoid the 
               influence of the upper states, which cannot be calculated 
               reliably in the truncated basis.
\end{thebibliography}
\end{document}